%% file: main.tex
\documentclass{article}
\usepackage{spconf}
\usepackage{cite}
\usepackage{amsmath,amssymb,amsfonts}
\usepackage{algorithm}
\usepackage{algpseudocode}
\usepackage{graphicx}
\usepackage{caption}
\usepackage{subcaption}
\usepackage{textcomp}
\usepackage{xcolor}
\usepackage{xcolor}

\usepackage{easyReview}

\DeclareMathOperator*{\amin}{argmin} 
\newcommand{\x}{\boldsymbol{x}}
\newcommand{\y}{\boldsymbol{y}}
\newcommand{\z}{\boldsymbol{z}}

\newcommand{\rl}{\boldsymbol{\ell}}
\newcommand{\hatl}{\boldsymbol{\hat{\ell}}}

\newcommand{\hy}{\boldsymbol{\hat{y}}}
\newcommand{\ar}{\mathcal{A}_\mathrm{R}}
\newcommand{\dar}{\Delta_{|\mathcal{A}_\mathrm{R}|}}
\newcommand{\U}{\textbf{U}}
\newcommand{\mH}{\mathcal{H}}

\title{Radar Anti-jamming Strategy Learning via Domain-knowledge Enhanced Online Convex Optimization\\
}

\name{Liangqi Liu\textsuperscript{$\star$}, Wenqiang Pu\textsuperscript{$\star$},Yingru Li\textsuperscript{$\star$}, Bo Jiu\textsuperscript{$\ddagger$}, Zhi-Quan Luo\textsuperscript{$\star$}\thanks{This work is supported by the National Nature Science Foundation of China (NSFC) under Grant 62101350. (\textit{Correspondence: Wenqiang Pu}). }}
\address{\textsuperscript{$\star$} Shenzhen Research Institute of Big Data, The Chinese University of Hong Kong, Shenzhen, China \\ 
\textsuperscript{$\ddagger$} National Key Laboratory of Radar Signal Processing, Xidian University, Xi’an, China}
\begin{document}
\ninept
\maketitle

\begin{abstract}
The dynamic competition between radar and jammer systems presents a significant challenge for modern Electronic Warfare (EW), as current active learning approaches still lack sample efficiency and fail to exploit jammer's characteristics. In this paper, the competition between a frequency agile radar and a Digital Radio Frequency Memory (DRFM)-based intelligent jammer is considered. We introduce an Online Convex Optimization (OCO) framework designed to illustrate this adversarial interaction. Notably, traditional OCO algorithms exhibit suboptimal sample efficiency due to the limited information obtained per round. To address the limitations, two refined algorithms are proposed, utilizing unbiased gradient estimators that leverage the unique attributes of the jammer system. Sub-linear theoretical results on both static regret and universal regret are provided, marking a significant improvement in OCO performance. Furthermore, simulation results reveal that the proposed algorithms outperform common OCO baselines, suggesting the potential for effective deployment in real-world scenarios.
\end{abstract}


\section{Introduction}\label{sec:intro}
\input{intro}

\section{System Model}\label{sec:sysmodel}
\input{signal_model}

\section{OCO-based Radar Strategy Design}
\input{oco}

\section{Proposed Algorithms}
\input{opponent}

\section{Experiments}
\input{experiment}

\section{Conclusion}
\input{conclusion}


\newpage
\bibliographystyle{IEEEtran}
\bibliography{ref}

\vspace{12pt}

\end{document}

%% file: intro.tex
The evolution of electronic warfare (EW) has introduced significant challenges for modern radar systems, especially due to the emergence of intelligent jammers~\cite{de2018introduction, adamy2001ew}. Among them, main lobe jamming poses a severe threat as jammers deliberately position themselves within the radar's main beam. Traditional signal processing techniques developed to detect and mitigate jamming signals often fall short, as they rely on specific assumptions that may not always be valid. For example, blind source separation assumes a non-zero angular separation between the radar and jammer\cite{geMainlobeJammingSuppression2018}, which is invalid in scenarios involving self-protection jammers~\cite{skolnik1970radar}.

To address the limitations of traditional jamming countermeasures, frequency agility has been adopted, enabling radar systems to implement adaptive frequency hopping strategies in the face of main lobe jamming~\cite{axelssonAnalysisRandomStep2007, zhouruixueCoherentSignalProcessing2015, bicaGeneralizedMulticarrierRadar2016}. Recently, the application of learning techniques has redefined anti-jamming efforts as sequential decision-making problems. In particular, reinforcement learning has been incorporated into the development of anti-jamming strategies, with notable research exploring the deployment of Deep Q-Networks (DQN)~\cite{zhengAirborneRadarAntiJamming2022, liRadarActiveAntagonism2021}. Game-theoretic methods, including two-person zero-sum games and extensive-form games with imperfect information, have been leveraged to enhance strategic anti-jamming performance~\cite{songMIMORadarJammer2012,gengRadarJammerIntelligent2023, liNeuralFictitiousSelfPlay2022}. Additionally, the multi-armed bandit framework has emerged as a novel approach for designing anti-jamming strategies~\cite{fangOnlineFrequencyAgileStrategy2022}. The field continues to advance with studies on subpulse-level frequency agile radar systems~\cite{liDeepQNetworkBased2019} and the development of adaptive power allocation models for jammers~\cite{gengRadarJammerIntelligent2023}. However, these active online learning techniques in anti-jamming strategy design either fall short in sample efficiency or struggle to model the dynamics of the jammer. 

To overcome these limitations, it is essential to effectively model the attributes of the jammer system. In this paper, we adopt the framework of Online Convex Optimization (OCO)~\cite{zinkevich2003online}, in which the jamming strategy can be naturally embedded in the gradient of the cost function. The online interaction information thus can be effectively utilized to estimate the gradient and a online mirror descent algorithm is developed. We show that for an arbitrary jammer, the developed algorithm achieves $O(\sqrt{N})$ static regret bound (comparing with the best fixed decision in hindsight), where $N$ is the total number of iterations. This improves the results of existing works~\cite{hazan2007logarithmic, shalev2012online, mokhtari2016online}. Further, by exploiting the knowledge that the jammer's decisions are predicated on past interactions, we devise a more efficient algorithm that achieves an $O(\sqrt{N})$ regret bound for universal regret. This sub-linear performance suggests the feasibility of reaching an optimal anti-jamming strategy and marks a significant improvement over classic OCO algorithms. The simulation results further demonstrates that proposed algorithms outperforms OCO baselines.

%% file: signal_model.tex
\subsection{Signal Model}
Consider a frequency-agile radar system that employs a subpulse-level frequency-agile waveform during transmission. The system works in a monopulse mode, a single pulse is composed of multiple sub-pulses, each operating at a unique carrier frequency. Let $\mathcal{F} = \{f_1, f_2, \ldots, f_L\}$ denote the set of available carrier frequencies, with each frequency $f_i$, $i = 2, 3, \ldots, L$, defined as $f_{i-1} + \Delta f$ and $\Delta f$ represents the frequency step size. Denoting $M$ as the total number of sub-pulses within one pulse, the transmitted frequency-agile signal of the $n$-th pulse $(n\in[N])$ at time $t$ is expressed as 
\begin{equation}\label{equ:signal_radar}
s_n(t) = \sum_{m=0}^{M-1}\text{rect}\left(\frac{t-mT_c}{T_c}\right)P_ma(t)\exp{(j2\pi f^{\textrm{R}}_m t)},
\end{equation}
where $T_c$ is the sub-pulse duration, $P_m$ represents the power allocated to the $m$-th subpulse, $a(t)$ is the complex envelope, and $\text{rect}(t)$ is the rectangle function defined as $\text{rect}(t)=1$ for $0\leq t\leq 1$ and $\text{rect}(t)=0$ otherwise. Parameter $f^{\textrm{R}}_m\in\mathcal{F}$ denotes the sub-carrier frequency for the $m$-th sub-pulse, which may vary among sub-pulses. 

The jamming system is equipped on the target, aims to mask the radar's reflected pulse echo by transmitting a noise-modulated signal. The distance between the radar and target/jammer is denoted as $R$,  and the jamming signal $j_n(t)$ can be described as
\begin{equation}\label{equ:signal_jammer}
j_n(t) =\sum_{m=0}^{M-1}\text{rect}\left(\frac{t-T_d - mT_c}{T_c} \right)b_m(t)\exp{(j2\pi f^\mathrm{J}_m t)}
\end{equation}
where $T_d=R/c$ ($c$ is the speed of light) is the propagation delay from the radar to the target/jammer, $b_m(t)$ represents the noise-modulated envelope, and $f^\textrm{J}_m$ corresponds to the carrier frequency of sub-pulse $m$. At the receiver, the radar captures a signal mixture consisting of the reflected radar pulse echo, the jamming signal, and additive noise. The received signal $r_n(t)$ is expressed as 
\begin{equation}\label{equ:signal_echo}
r_n(t) = s_n(t-T_d) + j_n(t - T_d) + w_n(t),
\end{equation}
where $w_n(t)$ denotes the additive white Gaussian noise.

\begin{figure}
    \centering
    \includegraphics[width=0.99\linewidth]{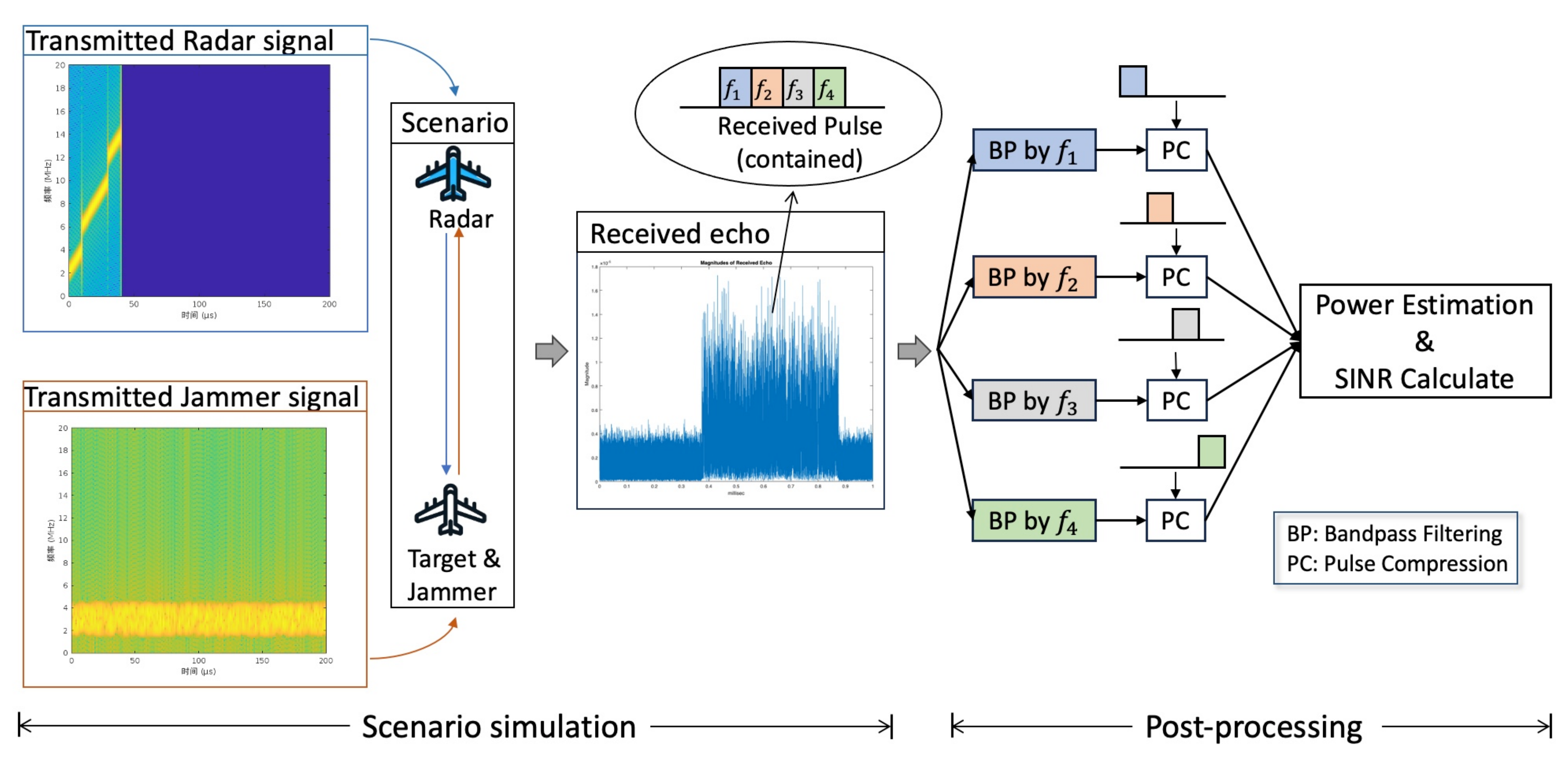}
    \caption{Illustration on the anti-jamming scenario.}
    \label{fig:model-scenario}
\vspace{-4mm}
\end{figure}

\subsection{Post-processing of Received Signal}
By implementing a suitable post-processing procedure on the mixed signal $r_n(t)$, it is possible to extract valuable information of the jamming signal and the radar echo. This procedure involves bandpass filtering to isolate the desired frequency band, followed by matched filtering designed to enhance the of subpulses in radar echo. The sequential steps of this procedure are depicted in the flowchart presented in Fig.~\ref{fig:model-scenario}. Crucially, this process facilitates the estimation of the power of each subpulses in radar echo, denoted as $P_\mathrm{R}(f^\mathrm{R}_m)$, the power of jamming signal in radar receiver $P_\mathrm{J}$, and the noise power $P_{n_0}$. Consequently, the Signal-to-Interference-plus-Noise Ratio (SINR) for each subpulse $m$ can be calculated, denoted as $\text{SINR}_m$,
\begin{equation}\label{equ:sinr}
\text{SINR}_m = \frac{P_\mathrm{R}(f^\mathrm{R}_m)}{P_{n_0}+P_\mathrm{J}\mathbf{1}(f^\mathrm{R}_m=f^\mathrm{J}_m)},
\end{equation}
where the indicator function $\mathbf{1}(f^\mathrm{R}_m=f^\mathrm{J}_m)$ determines whether the received echo is subject to jamming on sub-pulse $m$. $\text{SINR}_m$ indicates the quality of the $m$-th sub-pulse for target detection, which will be used as the utility function later.

%% file: oco.tex
\subsection{Online Optimization Formulation}
Based on the established signal models and post-processing procedure in Section~\ref{sec:sysmodel}, each pulse can be regarded as one round of interaction between the radar and jammer. Protocol~\ref{alg:protocol} presents the protocol for this type of iterative interaction. The radar system dynamically adjusts its transmitted signal $s_n(t)$ to improve the SINR, by utilizing information extracted from previous pulses. To systematically select parameters of $s_n(t)$ at each iteration for this online interaction process, we propose an online optimization framework from the radar's perspective. The goal is to enhance the radar's adaptive capabilities in response to jammer, ensuring that with each pulse, the radar's response is optimally adjusted to the evolving jamming strategies. Basic elements involved in this interaction process are given below.
\begin{algorithm}
\captionsetup{labelfont={bf},name={Protocol},labelsep=period}
\caption{Iterative anti-jamming procedure}
\label{alg:protocol}
\begin{algorithmic}
\For{pulse $n=1,2,\ldots,N$}
\State Radar transmits the signal $s_n(t)$;
\State Jammer detects $s_n(t)$ and immediately transmits $j_n(t)$;
\State Radar receives the echo $r_n(t)$ and calculate SINR from it.
\EndFor 
\end{algorithmic} 
\end{algorithm}  

\begin{itemize}
    \item \textbf{Radar's action} $a_n$: $a_n=(f^\mathrm{R}_{n,1},f^\mathrm{R}_{n,2},\ldots,f^\mathrm{R}_{n,M})\in\ar=\mathcal{F}^{M}$ represents the radar's choice of various sub-carrier frequencies at the $n$-th pulse.
\item \textbf{Radar's strategy} $\x_n$: The strategy from which $a_n$ is selected belongs to the probability simplex $\dar$.
\item \textbf{Cost function} $f_n$: The convex cost function takes a linear form as $f_n(\x)=\langle\rl_n,\x\rangle:\dar\rightarrow\mathbb{R}$.
\end{itemize}
Notably, for each pulse $n$, only $\rl(a_n)\in[0,1]$ is accessible, derived from~\eqref{equ:sinr} as
\begin{equation}\label{equ:cost}
    \rl(a_n)=\frac{c-\text{SINR}(a_n)}{c}\in[0,1],
\end{equation}
where $c$ is an SINR threshold that indicates high enough detection probability, $\text{SINR}(a_n)=\frac{1}{M}\sum_{m=1}^M \min(\text{SINR}_m(a_n), c)$, and $\text{SINR}_m(a_n)$ is the SINR of the $m$-th pulse associating with action $a_n$ (c.f. \eqref{equ:sinr}). For the $n$-th pulse, radar's goal is to find a strategy $\x_n$ that minimizes the cost function $f_n$ and the optimization problem is 
\begin{equation}\label{equ:obj}
\begin{aligned}
\min_{\x_n} \quad & f_n(\x_n) \\
\text{s.t.} \quad & \x_n\in\dar.
\end{aligned}
\end{equation}
Achieving the optimal solution $\x^*_n$ poses a significant challenge due to the dynamic and unknown nature of $f_n$. Consequently, for each pulse $n$, an implementable online algorithm $\mathcal{A}$ is employed. This algorithm uses the historical function values $(f_1,f_2,\ldots, f_{n-1})$, along with any other pertinent information, to generate the strategy $\x_n$. The action $a_n$ is then selected according to the probability distribution defined by $\x_n$. After executing the algorithm over $N$ pulses, the retrospectively best strategy $\x^*$ is defined as $\x^*=\arg\min_{\x\in\Delta_{|\mathcal{A}_R|}}\sum_{n=1}^N f_n(\x)$. A meaningful performance metric for the online algorithm $\mathcal{A}$ is the so-called static regret. This metric evaluates the cumulative cost incurred by algorithm $\mathcal{A}$ and compares it with the cost of $\x^*$,
\begin{equation}\label{equ:s-regret}
     \text{S-Regret}_N(\mathcal{A})=\sum_{n=1}^Nf_n(\x_n) - \sum_{n=1}^N f_n(\x^*).
\end{equation}
Algorithm $\mathcal{A}$ is said to be effective if its regret is sublinear as a function of $N$, i.e., $\text{S-Regret}_N(\mathcal{A})=o(N)$. This condition implies that on average, the algorithm performs as well as the best fixed strategy in hindsight.

\subsection{Online Mirror Descent}
Based on E.q.~\eqref{equ:s-regret}, the Online Gradient Descent (OGD)\cite{zinkevich2003online} can be used for updating $\x_n$, given as
\begin{equation}
    \x_{n+1}=\text{Proj}_{\dar}[\x_n-\eta_n\nabla f_n(\x_n)]
\end{equation} 
where $\text{Proj}{\dar}[\cdot]$ represents projection onto the feasible set $\dar$, and $\eta_n$ is the learning rate at iteration $n$. To ensure stability~\cite{hazanIntroductionOnlineConvex2023} of OGD, a regularization term named Bregman divergence $D_F(\x, \x_n)$\cite{banerjee2005clustering} is introduced, which is associated with a Legendre function $F$\cite{bauschke1997legendre} and its domain $\mathcal{D}$. Consequently, the update rule becomes
\begin{equation}\label{eq:omd}
\begin{cases}
&\boldsymbol{u}_{n+1}=\amin\limits_{\x\in\mathcal{D}}\ \eta_n\langle\x,\nabla f_n\rangle+D_F(\x, \x_n) \\
&\x_{n+1}=\amin\limits_{\x\in\dar}\ D_F(\x, \boldsymbol{u}_{n+1})     
\end{cases}
\end{equation}
This iterative process is known as Online Mirror Descent (OMD)~\cite{hazan2010extracting} and $F(x)=\sum(x_i\ln{x_i}-x_i)$ is commonly adopted ($x_i$ is the $i$-th element of $x$), which admits analytic expression for~\eqref{eq:omd}. Notice that during the online interaction, only $f_n(a_n)$ is observed, other values of $f_n(a), a\neq a_n$ are not available. This makes estimating $\nabla f_n$ be the key step of~\eqref{eq:omd}. A naive estimator is the importance weighted estimator (IWE)~\cite{flaxman2004online} given as 
\begin{equation} \label{equ:grad_exp3}
\hatl_n(a)=
\begin{cases}
\frac{f_n(a)}{\x_n(a)}, &a=a_n \\
0, & \text{otherwise}. 
\end{cases}    
\end{equation} 
Setting $\nabla f_n=\hatl_n$, the OMD-IWE algorithm in \eqref{eq:omd} is equivalent to a classical bandit algorithm known as Exp3~\cite{auer2002nonstochastic}, characterized by a sub-linear static regret bound of $\mathcal{O}(\sqrt{N|\ar|\log{|\ar|}})$. However, the IWE neglects the strategic nature of the interaction inherent in the online process, where the radar is essentially engaged in a game against a jammer. Critical insights into the jammer's behavior remain unutilized. By incorporating knowledge about the jammer's action space and strategy, we can develop more sophisticated and effective online algorithms.



%% file: opponent.tex
\subsection{Action Modeling}

By leveraging the characteristics of the jammer, its action and strategy are modeled as below. 
\begin{itemize}
    \item \textbf{Jammer's action} $b_n$: $b_n\in\mathcal{A}_\mathrm{J}=\mathcal{F}^M\cup\mathcal{B}$, where $\mathcal{B}$ represents  specific jamming action such repeater jamming \cite{wangMathematicPrinciplesInterruptedsampling2007, fengJammingWidebandRadar2017}.
\item \textbf{Jammer's  strategy} $\y_n$: $b_n$ is selected from its strategy $\y_n\in\Delta_{|\mathcal{A}_\mathrm{J}|}$, where $\Delta_{|\mathcal{A}_\mathrm{J}|}$ denotes the probability simplex.
\end{itemize}
The cost function $f_n$ thus can be related to the cost of a two-player matrix game. In particular consider a matrix game with cost matrix\footnote{$U[a,b], \forall a,b$ are calculated by E.q.~\eqref{equ:cost}.} $\U\in\mathbb{R}^{|\mathcal{A}_\mathrm{R}|\times|\mathcal{A}_\mathrm{J}|}$, at the $n$-th pulse, the radar player takes action $a_n\sim\x_n$, the jammer player takes action $b_n\sim\y_n$, then the following holds
\begin{equation}
    \mathbb{E}_{b_n\sim \y_n}[f_n(\x_n)] = \x_n^T\U\y_n := \phi(\x_n, \y_n),
\end{equation}

The above claim implies that an unbiased gradient estimator of $f_n(\x_n)$ can be attained as
\begin{equation}\label{eq:grad_est_a}
    \hatl_n = \U[:,b_n],
\end{equation}
where the unbiased property is due to 
$$\mathbb{E}[\hatl_n]=\U\times\mathbb{E}[\boldsymbol{b}_n]=\U\times\y_n=\rl_n$$
and $\boldsymbol{b}_n$ is the pure strategy of its action $b_n\sim\y_n$.

By incorporating the gradient estimator from \eqref{eq:grad_est_a} into the OMD~\eqref{eq:omd}, we introduce an enhanced algorithm referred to as OMD with Action Modeling Estimator (OMD-AME), which is elaborated in Algorithm \ref{alg:exp3em}. OMD-AME achieves a sub-linear static regret bound $\mathcal{O}(\sqrt{N\log{|\mathcal{A}|}})$, with the proof omitted here. This improved regret bound effectively removes the square-root dependence on the size of the action set $|\mathcal{A}|$ that characterizes the classical OMD-IWE (Exp3) algorithm. The key to this advancement lies in leveraging the inherent game structure of the cost matrix.
\begin{algorithm}
\setcounter{algorithm}{0}
\caption{Online Mirror Descent with Action Modeling Estimator (OMD-AME)}
\label{alg:exp3em}
\begin{algorithmic}
\State \textbf{Input}: $\eta>0$, $\x_1=(1/|\ar|)\times\mathbf{1}$
\For{pulse $n=1,2,\ldots,N$ \textbf{Radar}}  
\State Take action $a_n\sim\x_n$ and observe the echo $r_n(t)$;
\State \textbf{Gradient estimation} from $r_n(t)$ by E.q.~\eqref{eq:grad_est_a};
\State \textbf{Online mirror descent} with $\hatl_n$ by E.q.~\eqref{eq:omd};
\EndFor 
\end{algorithmic} 
\end{algorithm}
\vspace{-5mm}

\subsection{Opponent Modeling}
Modern jamming system usually contains a crucial subsystem named Digital Radio Frequency Memory (DRFM)~\cite{de2018introduction}, which records useful interaction histories that jammers can utilize to make informed decisions. By leveraging knowledge of the jamming strategy generation mechanism—specifically, the jamming strategy is a pre-defined \textit{fixed} decision rule based on a length-$k$ history $h_n = \{a_{n-1}, b_{n-1}, \ldots, a_{n-k}, b_{n-k}\}\in\mathcal{H}=\mathcal{A}_\mathrm{R}^k\times\mathcal{A}_\mathrm{J}^k$—a more effective online algorithm can be developed. Under this setting, jamming strategy $\y_n$ is re-expressed as 
$$\y_n =\pi\left(h_n\right),$$
where $\pi:\mathcal{H} \rightarrow\Delta_{|\mathcal{A}_\mathrm{J}|}$ is defined as the mapping from history space to the jammer's action space. Thus, the gradient estimator $\hatl_n$ can be re-formulated by associating the length-$k$ history $h_n$,
\begin{equation}\label{equ:grad-fixrule}
    \hatl_n = \U\hy_n,
\end{equation}
where $\hy_n=\pi_n(h_n)$  represents the estimate of the decision rule $\pi$ at pulse $n$. A natural choice for $\pi_n$ is the maximum likelihood estimator (MLE), which calculates the frequency of each action $b \in \mathcal{A}_\mathrm{J}$, conditioned on a length-$k$ history $h \in \mathcal{H}$, across the entire interaction history $(a_1, b_1, \ldots, a_{n-1}, b_{n-1})$. Combining gradient estimator \eqref{equ:grad-fixrule} with OMD~\eqref{eq:omd}, we attain OMD with Opponent Modeling Estimator (OMD-OM), elaborated in Algorithm~\ref{alg:singlemodel}.

Notably, for OMD-OME, the static regret defined in~\eqref{equ:s-regret} may tend to negative as it only compares with the best strategy $\x^*$ in hindsight. This comparator is weak since OMD-OME exploits the jammer strategy information. To address this limitation, we introduce a broader performance metric termed \emph{universal} regret, which aligns more closely with the objective function described in \eqref{equ:obj}.
\begin{equation}\label{equ:uni-regret}
    \text{U-Regret}_N(\mathcal{A})=\sum_{n=1}^Nf_n(\x_n) - \min_{\z_1,\ldots,\z_N}\sum_{n=1}^N f_n(\z_n)
\end{equation}
where $(\z_1,\ldots,\z_N)$ is a comparator sequence. Furthermore, if $\z_1=\ldots=\z_N=\x=\amin\limits_{\x} \sum\limits_{n=1}^N f_n(\x)$, it reverts to static regret. Typically, achieving sub-linear universal regret is unattainable due to the unknown dynamics of the environment. Nevertheless, OMD-OME, which exploits the information of jamming strategy generation, utilizes a history-dependent predictor to capture the dynamic nature of the jamming strategy. By doing so, it can attain a sub-linear universal regret bound, expressed as $\mathcal{O}(\sqrt{N|\mathcal{H}|\log{|\mathcal{A}_\mathrm{R}|}})$ with the proof omitted here. This bound represents a significant improvement over the sub-linear static regret bounds obtained with traditional OMD-IWE (Exp3) and improved OMD-AME algorithm, offering a mechanism to adapt to the dynamics of jamming environments effectively.
\vspace{-1mm}

\begin{algorithm}
\caption{Online Mirror Descent with Opponent Modeling Estimator (OMD-OME)}
\label{alg:singlemodel}
\begin{algorithmic}
\State \textbf{Input}: $\x_1$, length-$k$ history $\mH$. 
\For{pulse $n=1,2,\ldots,N$ \textbf{Radar}} 
\State Take action $a_n\sim\x_n$ and observe the echo $r_n(t)$;
\State Update the history $\mH$ and obtain $\hy_n$ by MLE;
\State \textbf{Gradient estimation} by E.q.~\eqref{equ:grad-fixrule};
\State \textbf{Online mirror descent} with $\hatl_n$ by E.q.~\eqref{eq:omd};

\EndFor 
\end{algorithmic}
\end{algorithm}
\vspace{-5mm}

%% file: experiment.tex
\begin{figure}[t] 
\centering

\begin{subfigure}[b]{0.49\columnwidth}
    \centering
    \includegraphics[width=\textwidth]{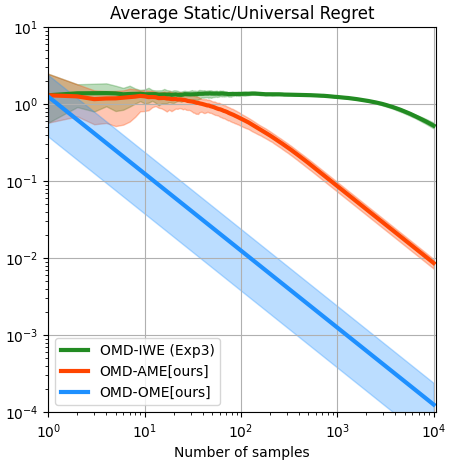}
    \caption{Stationary Environment}
    \label{fig:sta}
\end{subfigure}
\hfill 
\begin{subfigure}[b]{0.49\columnwidth}
    \centering
    \includegraphics[width=\textwidth]{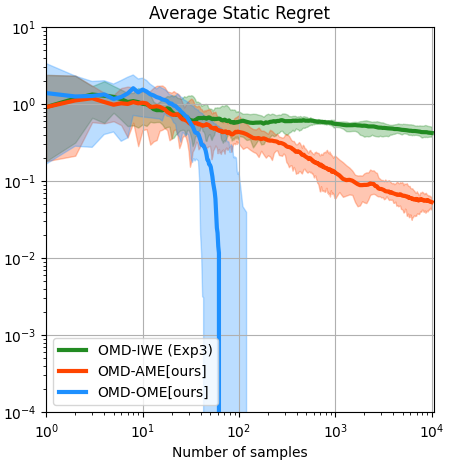}
    \caption{Non-stationary Environment}
    \label{fig:nonsta-1}
\end{subfigure}


\begin{subfigure}[b]{0.49\columnwidth}
    \centering
    \includegraphics[width=\textwidth]{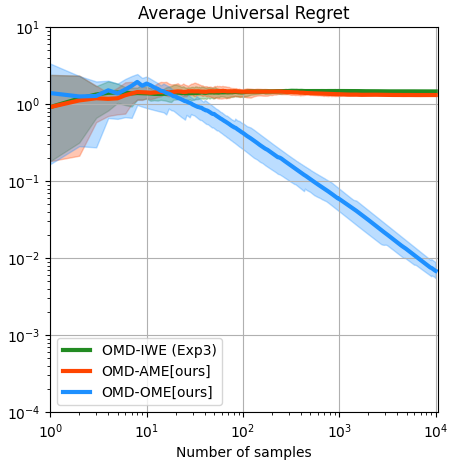}
    \caption{Non-stationary Environment}
    \label{fig:nonsta-2}
\end{subfigure}
\hfill
\begin{subfigure}[b]{0.49\columnwidth}
    \centering
    \includegraphics[width=0.95\textwidth]{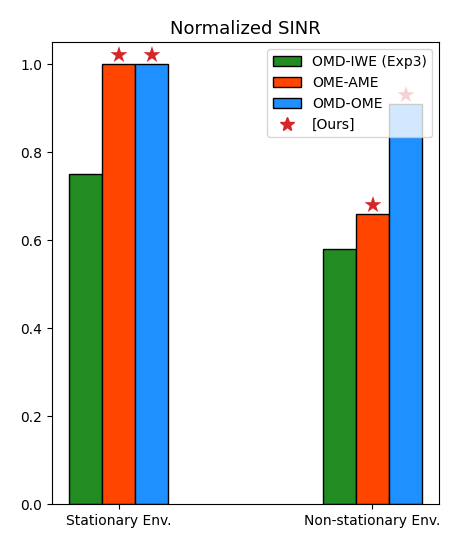}
    \caption{SINR Comparison}
    \label{fig:hist}
\end{subfigure}
\caption{Comparative analysis of baseline OMD-IWE with proposed OMD-AME and OMD-OME methods, focusing on regret and SINR performance. (a) Regret comparison in a stationary env., where static and universal regrets coincide. (b) Static regret comparison in a non-stationary env. (c) Universal regret comparison in a non-stationary env. (d) SINR values after $1000$ interactions across both env.}
\label{fig:expe}
\vspace{-4.5mm}
\end{figure}

\begin{table}[t]
\centering
\begin{tabular}{|c|c|}
\hline
Parameter             & Value      \\ \hline
Radar antenna gain                & 30dB  \\
Radar transmitted power $P_R$                   & 10KW       \\
Jammer transmitted power $P_J$                   & 1KW        \\
\# of sub-pulses        & 4          \\
Sub-pulse width         & 3$\mu$s   \\
PRF                     & 1000Hz       \\
Carrier frequency       & 10GHz     \\
Distance                & 100KM      \\ \hline
\end{tabular}
\caption{Parameter setting for FA radar and jammer.}
\label{tab:para}
\vspace{-5mm}
\end{table}
This section evaluates the efficiency of the developed OMD-AME and OMD-OME algorithms through simulation experiments. Detailed settings for the radar and jammer systems are provided in Table \ref{tab:para}. The static regret \eqref{equ:s-regret} and universal regret~\eqref{equ:uni-regret} are employed as the performance metrics. Two different jamming strategies were simulated: a stationary strategy where $\y_n=\y$ remains constant for all iterations within the simplex $\Delta_{|\mathcal{A}_J|}$, and a non-stationary strategy that varies in response to recent length-$3$ history, denoted as $\y_n=\pi(h_n), h_n\in\mH$. Specifically, $\y_n(b_1)=0.7$ and $\y_n(b_2)=0.3$ where $b_1, b_2$ are related with the two most common frequencies in $h_n$. The OMD-IWE (Exp3) algorithm serves as the baseline for comparisons on static regret and universal regret. For each algorithm, 500 independent trials are conducted, and the shaded areas in Fig.\ref{fig:expe} represent the confidence intervals of these trials.

Fig.~\ref{fig:sta} compares the average static/universal regret under a stationary jamming strategy. In the stationary environment, the two regrets are identical. The results demonstrate that all evaluated algorithms achieve sub-linear regret. Notably, OMD-OME and OMD-AME outperform the baseline OMD-IWE (Exp3) algorithm, significantly reducing the number of required samples.

In Fig.~\ref{fig:nonsta-1} and~\ref{fig:nonsta-2}, static and universal regrets under a non-stationary jamming strategy are displayed. With regard to static regret in Fig.~\ref{fig:nonsta-1}, OMD-AME and OMD-OME exhibit more rapid declines compared to the baseline OMD-IWE (Exp3), indicating a faster convergence rate. Notably, OMD-OME demonstrates a substantial reduction in static regret, highlighting its smaller cumulative cost compared with the optimal static strategy. For universal regret, only OMD-OME achieves a sub-linear regret, showcasing its impressive capability to adapt to non-stationary environments.

Finally, Fig.~\ref{fig:hist} compares the normalized SINR values received by the radar system using different online algorithms after $1000$ interactions. This comparison directly demonstrates the effectiveness of each method within a limited sample size. The proposed OMD-AME and OMD-OME achieve superior SINR in both stationary and non-stationary, showcasing their effectiveness and sample efficiency.

\noindent\textbf{Limitations} Our study does have certain limitations that should be addressed in future work. A key challenge arises from the practical application scenarios where the set of available frequencies, $\mathcal{F}$ in \eqref{equ:signal_radar}, is typically large. The size of the action set $\ar$ thus increases exponentially, significantly impairing sample efficiency across all methodologies. Specifically, in our proposed method, although the opponent modeling approach has substantially enhanced sample efficiency, the large action set still demands a more efficient estimation approach for the 'fixed rule' $\pi$, beyond the traditional MLE method.


\vspace{-2mm}


%% file: conclusion.tex
\vspace{-2mm}
This paper formulates the competition between subpulse-level frequency-agile radar and main-lobe intelligent jammer as an online convex optimization problem. Through careful modelings of the jamming strategies, we have developed algorithms that outperform conventional OCO benchmarks. Sub-linear static and universal regret bounds are provided, and numerical simulations demonstrates a significant enhancement in sample efficiency.